# ORIGIN OF WEAK-LINK BEHAVIOR OF GRAIN BOUNDARIES IN SUPERCONDUCTING CUPRATES AND PNICTIDES

GUY DEUTSCHER

School of Physics and Astronomy, Tel Aviv University, Tel Aviv 69978, Israel

**ABSTRACT**

*Superconducting cuprates and pnictides composed of $CuO_2$ or $AsFe$ planes respectively with intercalated insulating layers, are at the crossroads of three families of crystalline solids: metals, doped Mott insulators, and ferroelectrics. In the latter atomic displacements play a key role. Both the metallic and the doped insulator approaches to high temperature superconductivity are essentially electronic ones and do not directly involve the lattice. By contrast, in a recently proposed Bond Contraction Pairing (BCP) model, contraction of in-plane Cu-O (or As-Fe) bonds plays an essential role in the pairing mechanism. Here we apply it to low angle grain boundaries and show that their reduced critical current is due to tensile deformation generated by dislocations. The model also explains why interface misfit dislocations, which can result in a dead layer in the case of ferro-electrics, may improve vortex pinning in the cuprates.*

The low critical current across grain boundaries is arguably the most serious problem that stands in the way of manufacturing high-critical current, mile-long HTS wires. Typically, grain boundaries reduce the critical current density of optimally doped polycrystalline YBa$_2$Cu$_3$O$_{7+x}$ (YBCO) at 77K from an intra-grain value $j_c$ of several $1.10^6$ A/cm$^2$ down to a few 100A/cm$^2$. Studies of the intergrain critical current density $j_{cb}$ in artificial grain boundaries in thin films grown on SrTiO$_3$ bi-crystals have shown that its decrease as a function of the boundary angle is universal amongst all cuprates.[1] Even at angles lower than 10 degrees it is already substantial.[2] Recently, it has also been found that $j_{cb}$ in the pnictides behaves similarly to that in the cuprates, which underlines that its angular dependence is a fundamental property of a large class of high temperature superconductors.[3] As noted by Hilgenkampf and Mannhart,[1] understanding the grain boundary problem requires having a pairing model for High Tc. Since the lattice does not play an important role either in the metallic or in the doped Mott insulator approaches, it is doubtful that they can provide a suitable basis for an explanation, beyond the general remark that the HTS short coherence length makes them sensitive to small scale disorder.[4]

We will argue here that the origin of the low $j_{cb}$ is the tensile strain generated by dislocations located along the boundary. In the BCP model the formation of pairs, which is a precursor to long range superconducting order, is necessarily accompanied by a contraction of in-plane Cu-O (or As-Fe) bonds involving overlap of $d$ and $p$ orbitals.[5] We will show that a tensile strain larger than about 1% is sufficient to quench this pairing mechanism, and that such strain is present up to distances of several nanometers from dislocation cores where it generates non-pairing regions. When the boundary angle is larger than 5 to 10 degrees these regions overlap and turn the entire grain boundary into a weak junction. We also show that the same strain effect that is detrimental in the case of grain boundaries can enhance vortex pinning near misfit dislocations at interfaces with a substrate having a lattice parameter larger than that of the superconductor, and similarly in the vicinity of embedded nano-particles.

We first consider strain fields generated in the vicinity of a low angle grain boundary where edge dislocations are located at a distance L from each other. For a tilt boundary defined by a rotation angle θ around an axis perpendicular to the $CuO_2$ planes:

$$L = \frac{a}{2\sin\theta/2} \quad (1)$$

where *a* is one of the two lattice parameters of the $CuO_2$ (ab) plane (the slight difference between the values of the *a* and *b* lattice parameters in the orthorhombic superconducting phase of the cuprates is not important here). According to the theory of elasticity, the strain field for an edge dislocation (reasonably far from the core) is given by:

$$\varepsilon_{xx} = -c\frac{y(3x^2 + y^2)}{(x^2 + y^2)^2} \quad (2)$$

where we have taken the *x* axis parallel to the Burger's vector of the dislocation. The origin is taken at the location of the added half atomic plane. Here *c* expressed in % is a materials property, and coordinates are normalized by the lattice parameter.

The strain across the grain boundary is a tensile one on the side where an atomic plane is missing, and a compressive one on the side where it is added. In the case of the cuprates, the compressive strain is quickly released possibly by removal of an oxygen atom linking two Cu atoms, so that the strain along the boundary is found to be mostly a tensile one.[6] We have reproduced in the insert of Fig.1 the schematic representation of an atomic resolution picture of a 7 degrees angle grain boundary by Song et al.[7] A tensile distortion of the lattice in the direction perpendicular to the boundary is clearly seen below the dislocation core. As shown Fig.1 the strain $\varepsilon_{xx}(y)$ can be fitted reasonably well to Eq.2. It reaches a 1% level between the 5$^{th}$ and the 10$^{th}$ row below the dislocation (c ≈ 5 to 10%).

In the BCP model, the pair breaking energy is given by:

$$2\Delta = 4\frac{(t_{CuO})^2}{U} - 8t_0 \tag{3}$$

where we have used for definiteness notation relevant for the case of the cuprates. Here the first term on the r.h.s. is the energy gained by two doped carriers when they localize on opposite sides of an Oxygen atom forming a singlet pair,[5] and the second term is the energy gained by releasing two electrons at the bottom of the conduction band (or at the top of the valence bad if these are holes), i.e. the band width. The transfer integral $t_{CuO}$ between a $d$ orbital state on a Cu atom and a $p$ orbital state on the neighboring oxygen is that in the presence of an added carrier to the pristine AF compound, and $U$ is the on-site Coulomb repulsion. For pairs to form, the difference between the two terms of the r.h.s. must be positive.

Pair formation results from a delicate balance between two large terms that can be easily upset. In the BCP model, $\Delta$ is the pseudo-gap, known to be of the order of 50 meV. Both terms on the r.h.s. are of the order of one eV, larger than the pseudo-gap by one order of magnitude. Pairing is possible through Cu-O bond contraction because the transfer integral $t_{CuO}$ increases quickly when the Cu-O bond length is shortened, as it varies as the 5$^{th}$ power of that length. *Vice versa* a small elongation of that bond is sufficient to prevent pair formation. Quantitatively, a tensile strain of the order of 1% is sufficient to suppress pairing since the difference between the two terms of the r.h.s. of Eq.3 is of about 10%. Since strains of that order are reached as we have seen up to the fifth to tenth row below a dislocation core along the boundary, pairing is suppressed up to that distance. If $L$ is smaller than 5 to 10 atomic lattice parameters, pairing will be suppressed along the entire boundary. This distance is reached when the boundary angle is larger than 5 to 10 degrees. This prediction of the BCP model is in full agreement with the data of Heinig et al. which show that low angle values $j_{cb}$ extrapolate down to zero at an angle of about 10 degrees.[2] It is also in agreement with the recent data of Lee et al. who have shown that in the pnictides 6 degrees angle grain boundaries have already a reduced $j_{cb}$ value.[3] We emphasize that this angle is much lower than that at which dislocation cores overlap (about 30 degrees), an experimental result that was so far not well understood.

As long as regions where strain is higher than 1% do not overlap ($\theta < \theta_c \approx$ 5 to 10 degrees) the critical current across the boundary is dominated by the fraction of the boundary where pairing is not affected by strain. When they do ($\theta > \theta_c$), $j_{cb}$ is set by the profile of the order parameter *across* the non-pairing region. As one approaches the boundary the strain increases progressively as described by Eq.2 but, since the energy gained by pairing decreases as the 10$^{th}$ power of the strain, we model the transition from the pairing S to the non-pairing N region as one where the pairing potential V falls down abruptly to zero at a distance $x_c$ from the boundary (Fig.2). The ratio of the critical current across the boundary to the bulk one is then given by:

$$\frac{j_{cb}}{j_c} \cong \left(\frac{T_c - T}{T_c}\right)^{1/2} \exp - 2Kx_c$$

(4)

where $K$ is the inverse of the decay length of the order parameter in the non-pairing region, here of the order of the coherence length in S since S and N are identical, except for the value of the pairing potential.[8] This expression includes the effect of the temperature dependence of the order parameter at the boundary between the S and N regions, which results in $j_{cb}$ varying as $(T_c-T)^2$ as observed,[9] and that of the exponential decrease of the order parameter inside the non-pairing region. From Eq.2 a contour of equal strain surrounds a region whose total thickness across the boundary is about twice its extension along it. Defining the contour of the non-pairing region as that inside which the strain is larger than 1%, we call its extension $y_c$ along the boundary and $x_c \approx 2y_c$ across it. Taking T=77K, $y_c$ = 2.2nm from the extrapolated angle of 10 degrees measured by Heinig et al.[2] (corresponding to $c$ = 0.06), and $K^{-1}$ =1.5 nm, we obtain from Eq.4 a reduction factor of 50 when overlaps occurs, in good agreement with their data for $j_b$ at θ=10 degrees. The increased critical current obtained by Ca overdoping of the boundary [10] may be due to a longer decay length.

While tensile strains due to dislocations are detrimental in the case of grain boundaries because they turn the entire grain boundary plane into a non-pairing region when θ> θ$_c$ , they can be beneficial in the case of misfit dislocations resulting from a  lattice mismatch at an interface between a single crystal substrate and a cuprate film grown by hetero-epitaxy.[11] In the framework of the BCP model, tensile strain generated by misfit dislocations results in the  formation of non-pairing columnar regions having a diameter of the same order as the thickness of the non-pairing region across a grain boundary. The distance $d$ between misfit dislocations is equal to $a/f$ where $f$ is the lattice mismatch.. For an YBCO film grown on a $SrTiO_3$ substrate the mismatch is of about 2%, the distance between dislocations about 20 nm, hence $d \approx 10\, x_c$. This is favorable since the non-pairing regions do not overlap, and at the same time $d$ is sufficiently small to allow efficient vortex pinning in fields of several Tesla. Partial destruction of superconductivity in the cuprates due to misfit dislocations (here beneficial) is reminiscent of their effect on ferro-electric thin films.[12] It was shown that in the case of a lattice mismatch of several %, ferro-electricity is actually quenched near the interface, resulting in a "dead layer". No such destruction of ferro-electricity was observed in the case of a lattice mismatch of 0.1% or less, therefore the destruction at large lattice mismatch is clearly related to the formation of a sufficiently dense array of misfit dislocations

Misfit dislocations or other columnar defects are very effective in pinning vortices for fields oriented perpendicular to the interface.[11,13] However what is ideally desired for practical applications such as magnets is a completely isotropic critical current since in a coil there are magnetic fields of any orientation. It has recently been shown that this objective can be achieved when incoherent nanoscale insulating particles are embedded in the conductor.[14] Strain distribution in that case is more complicated than what it is for edge and misfit dislocations. However one may retain the general idea that a non-pairing shell having a thickness of a few nanometers surrounds these particles. It is this non-pairing shell that pins the vortices rather than the particles themselves which are just insulating bodies interrupting vortices. The strain distribution is not known in the case of embedded particles of different kinds, but the average micro-strain can been determined by an analysis of the diffraction lines broadening. It was found that Tc decreases when the micro-strain is larger than 1%, and does not when it is lower than 0.5%.[14] These results are in good agreement with the BCP model since superconductivity must be quenched in a substantial fraction of the sample if the average strain is larger than 1%.

In conclusion, based on a model where pairing requires a contraction of Cu-O bonds,[5] we have proposed that the tensile strain field surrounding edge dislocations destroys pairing in a region extending up to a few coherence lengths from the dislocation core. This strain effect turns grain boundaries into weak junctions as soon as the boundary angle is large enough for these regions to overlap along it, which according to our estimate occurs when it is btween 5 and 10 degrees, well before dislocation cores overlap as observed.[8] On the other hand, the same model explains the enhancement of the critical current density near interfaces where misfit dislocations induce tensile strain in the superconductor and in the vicinity of embedded nano-particles, because in these cases the strain effect creates non overlapping non-pairing regions that are favorable for vortex pinning. Our prediction of a broad non-pairing but conducting region extending up to several nano-meters across a grain boundary can be checked by Scanning Tunneling Microscopy without requiring atomic resolution, thus offering a test of the Bond Contraction Pairing model.

Inspiring discussions with David Larbalestier and Xavier Obradors are gratefully acknowledged. This work was supported in part by a grant from the Israel Ministry of Infrastructure

guyde@tau.ac.il, phone 00 972 3 6408205

**Figure captions:**

**Figure 1**

Below the dislocation core as shown in the insert, where he expansion of the lattice below the dislocation core in the direction perpendicular to the boundary is clearly seen, the strain $\varepsilon_{xx}(y)$ can be fitted to the 1/y dependence predicted by Eq. 2 for x=0. The 1% strain level is reached between the 5$^{th}$ and the 10$^{th}$ row below the dislocation. The insert is reproduced from Ref. 7.

**Figure 2.**

The tensile strain $\varepsilon_{xx}$, pairing potential V and order parameter $\psi$ across a grain boundary in the vicinity of a dislocation core. Pairing is suppressed by strain in a region **N** whose thickness across the boundary $2x_c$ is a few times larger than the coherence length. A finite order parameter is induced in **N** by the proximity effect with the banks **S**.

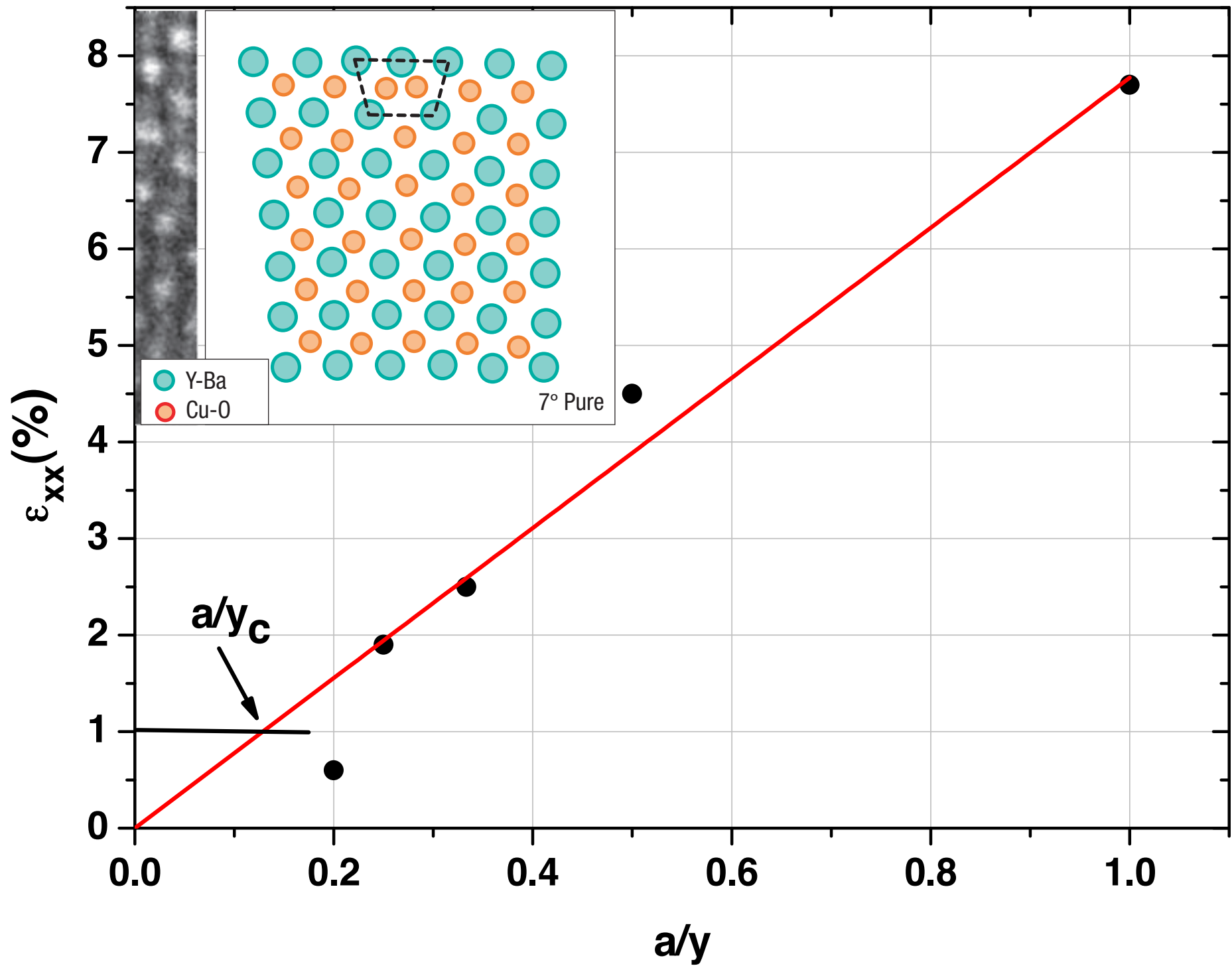

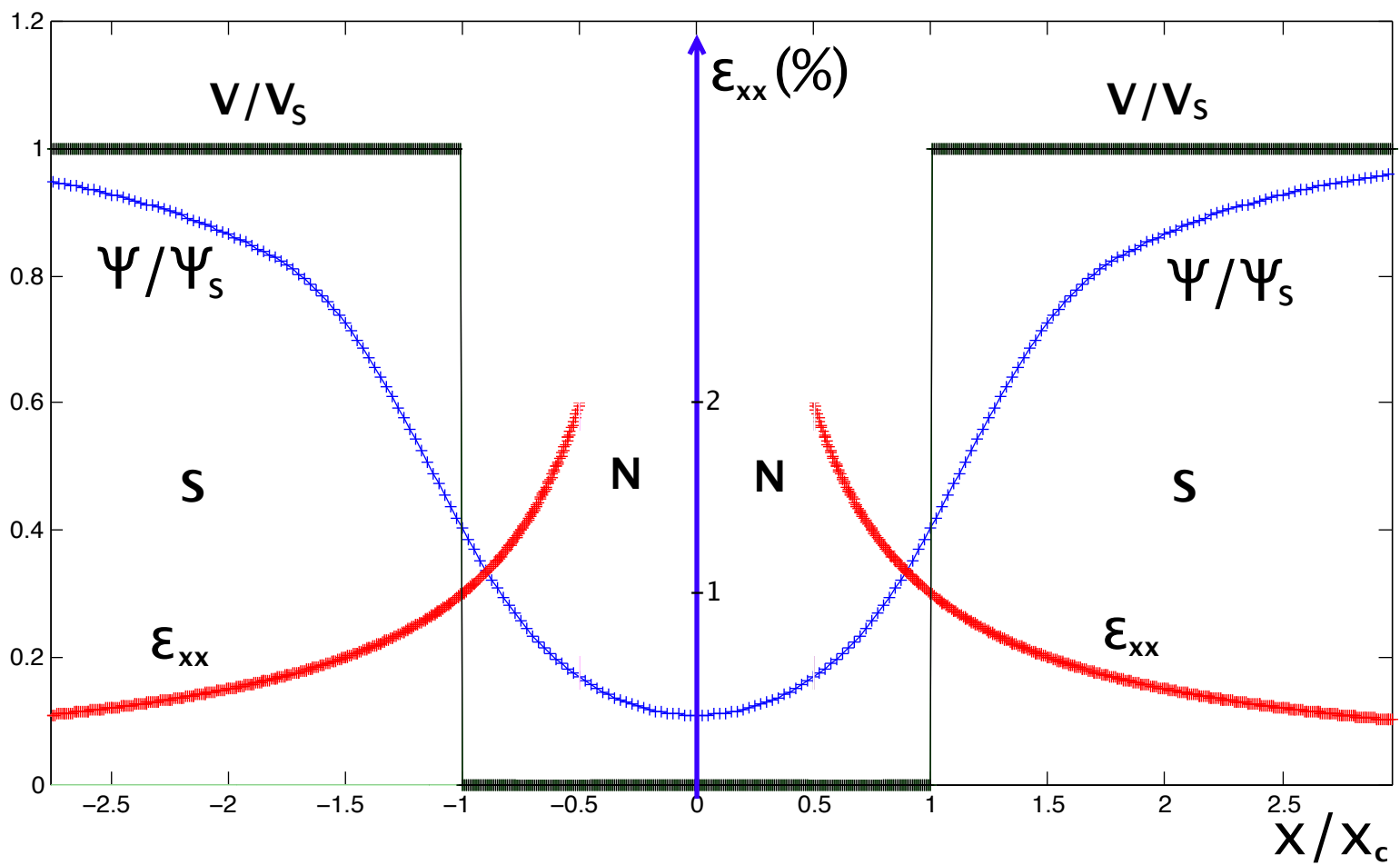